\documentclass[prl,twocolumn,final,showpacs,showkeys,nobibnotes,titlepage,twoside,10pt]{revtex4-1} 
\usepackage{amsmath}
\usepackage{graphicx}
\usepackage{amsfonts}
\usepackage{amssymb}
\usepackage{subfigure}
\usepackage{float}
\usepackage{color}
\usepackage{standalone}

\begin{document}

\title{Long-lived neighbors determine the rheological response of glasses}
\author{M. Laurati$^{1,2}$, P. Ma{\ss}hoff$^{1}$, K. J. Mutch$^{1}$, S. U. Egelhaaf$^{1}$, and A. Zaccone$^{3}$}
\affiliation{${}^1$Condensed Matter Physics Laboratory, Heinrich Heine University, 40225 D\"{u}sseldorf, Germany}
\affiliation{${}^2$Divisi\'on de Ciencias e Ingenier\'ia, Universidad de Guanajuato, Le\'on 37150, Mexico}
\affiliation{${}^3$Department of Chemical Engineering and Biotechnology, and Cavendish Laboratory, University of Cambridge, Cambridge CB2 3RA, UK}
\date{\today}

\begin{abstract}

Glasses exhibit a liquid-like structure but a solid-like rheological response with plastic deformations only occurring beyond yielding. Thus, predicting the rheological behavior from the microscopic structure is difficult, but important for materials science. Here, we consider colloidal suspensions and propose to supplement the static structural information with the local dynamics, namely the rearrangement and breaking of the cage of neighbors. This is quantified by the mean squared nonaffine displacement and the number of particles that remain nearest neighbors for a long time, i.e.~long-lived neighbors, respectively. Both quantities are followed under shear using confocal microscopy and are the basis to calculate the affine and nonaffine contributions to the elastic stress, which is complemented by the viscous stress to give the total stress. During start-up of shear, the model predicts three transient regimes that result from the interplay of affine, nonaffine and viscous contributions. Our prediction quantitatively agrees with rheological data and their dependencies on volume fraction and shear rate. 

\end{abstract}


\maketitle


Although glasses behave like solids, they have a liquid-like structure. Liquids and glasses exhibit similar static (time averaged) structural properties, such as the radial distribution function, and two- and four-point correlation functions \cite{Goetze}. These quantities thus cannot be directly responsible for the mechanical properties of glasses. In contrast, the diffusive dynamics of liquids distinguishes them from glasses, which show arrested dynamics and caging \cite{Goetze, Pusey87}. Under shear, the cage of nearest neighbors is deformed and becomes anisotropic \cite{Koumakis2012, mutch2013, Koumakis2016}, which has been studied by experiments and mode coupling theory (MCT) and linked to enhanced superdiffusive dynamics \cite{Zausch2008,Koumakis2012,mutch2013, Koumakis2016,Sentjabrskaja2014} and negative stress correlations \cite{Zausch2008,Amann2013}, and might lead to yielding \cite{Osuji2013,Benzi2014}. This suggests that, to understand the mechanical response of glasses, the effects of shear on the structure and dynamics need to be considered for individual particles as well as their neighbors.

Symmetry is also important. A centrosymmetrical crystal deforms affinely under shear with its free energy increasing quadratically with applied strain \cite{Alexander98,Lemaitre06}.
However, in a glass, the particles are not local centers of symmetry. Hence, upon deformation, the forces on each particle do not balance and result in an additional net force. This is relaxed through nonaffine motions that lower the free energy and the shear modulus \cite{Zaccone2011}. Thus, shear induces both affine and nonaffine displacements~\cite{Alexander98}, and their quantitative description is a prerequisite to predict the rheological response of glasses.

As affine and nonaffine displacements cannot be distinguished based on structural properties alone, dynamical features must be included. The importance of cages for the glass transition suggests to consider these entities and their rearrangements. Previous approaches, like MCT, considered cage rearrangements on a mean-field level in terms of shear-induced (mean) cage anisotropy \cite{Zausch2008,Koumakis2012,mutch2013, Koumakis2016,Sentjabrskaja2014,Amann2013}. In contrast, using confocal microscopy we explicitly consider strain-induced rearrangements on a single-particle level, in particular of the nearest neighbors. Elasticity is found to be conferred by particles that remain nearest neighbors for a long time, i.e.~`long-lived neighbors'. Furthermore, instead of single-particle dynamics, we determine the dynamics of particles with respect to their nearest neighbors, quantified by their mean squared local nonaffine displacement \cite{falk_langer, chikkadi2012}.
Based on these experimentally accessible microscopic parameters, we calculate, from first principles, the total stress, which includes the affine and nonaffine contributions to the elastic stress as well as the viscous stress. 

To test and illustrate this approach, we investigate a colloidal model system and its shear-induced affine and nonaffine displacements during start-up of shear. This is a standard rheological test which is routinely applied to a broad range of systems, including colloidal glasses, gels and polymers \cite{Koumakis2012, mutch2013, Koumakis2016, Sentjabrskaja2014, Zausch2008, Laurati2012, Divoux2011, Varnik2004, Santos2013, Park2013, Amann2013, Colombo2014, Hasan1995, Koumakis2012b, Petekidis2004, Fuchs2005, Varnik2006}. Nevertheless, its link to the single particle level is only starting to be explored experimentally \cite{Koumakis2012, mutch2013, Koumakis2016, Sentjabrskaja2014, Zausch2008, Laurati2012, Divoux2011}.


We investigated dispersions of polymethylmethacrylate (PMMA) hard-sphere like particles \cite{Royall2013} with volume fractions $0.565 \le \phi \le 0.600$ \cite{Poon2012}, i.e.~around the glass transition~\cite{Pusey87}. The samples contained either small spheres (radius $R_1=150$~nm, polydispersity $\sigma_{\text{R,1}} \approx 12\%$) or large, fluorescently labelled spheres ($R_2=780$~nm, $\sigma_{\text{R,2}} \approx 6\%$).
Start-up shear experiments were performed using a stress-controlled rheometer (TA Instruments, DHR3) for the small particles, and a strain-controlled rheometer (TA Instruments, ARES-G2) for the large particles, with cone-plate geometries. 
For the microscopy experiments, we used a home-built shear cell (described in \cite{Laurati2012}).
Loading and history effects were reduced by a rejuvenation procedure before starting each measurement.
Image stacks were acquired using a confocal unit (Visitech, VT-Eye) mounted on an inverted microscope (Nikon, Ti-U).
The stacks of $512{\times}512{\times}50$ pixels or $51{\times}51{\times}10\,\mu$m$^3$ contain $\sim$8500 particles and were acquired in $\Delta t = 1.83$~s.
Further experimental details are in the SM \cite{SM}. 
From one series of confocal images, particle coordinates and trajectories were extracted using standard routines and then refined \cite{crocker96, Jenkins2008}.


The structure and dynamics of the large particles with $\phi = 0.565$ were investigated by confocal microscopy.
The mean number of nearest neighbors in the quiescent sample, $n_{\text{tot}}(\gamma{=}0) = 11.6$,
was determined from the integral of the first peak of the radial distribution function $g(r)$, taken up to the first minimum at $r = 2.6\,R$.

\begin{figure}
\centering
\includegraphics[width=0.75\columnwidth]{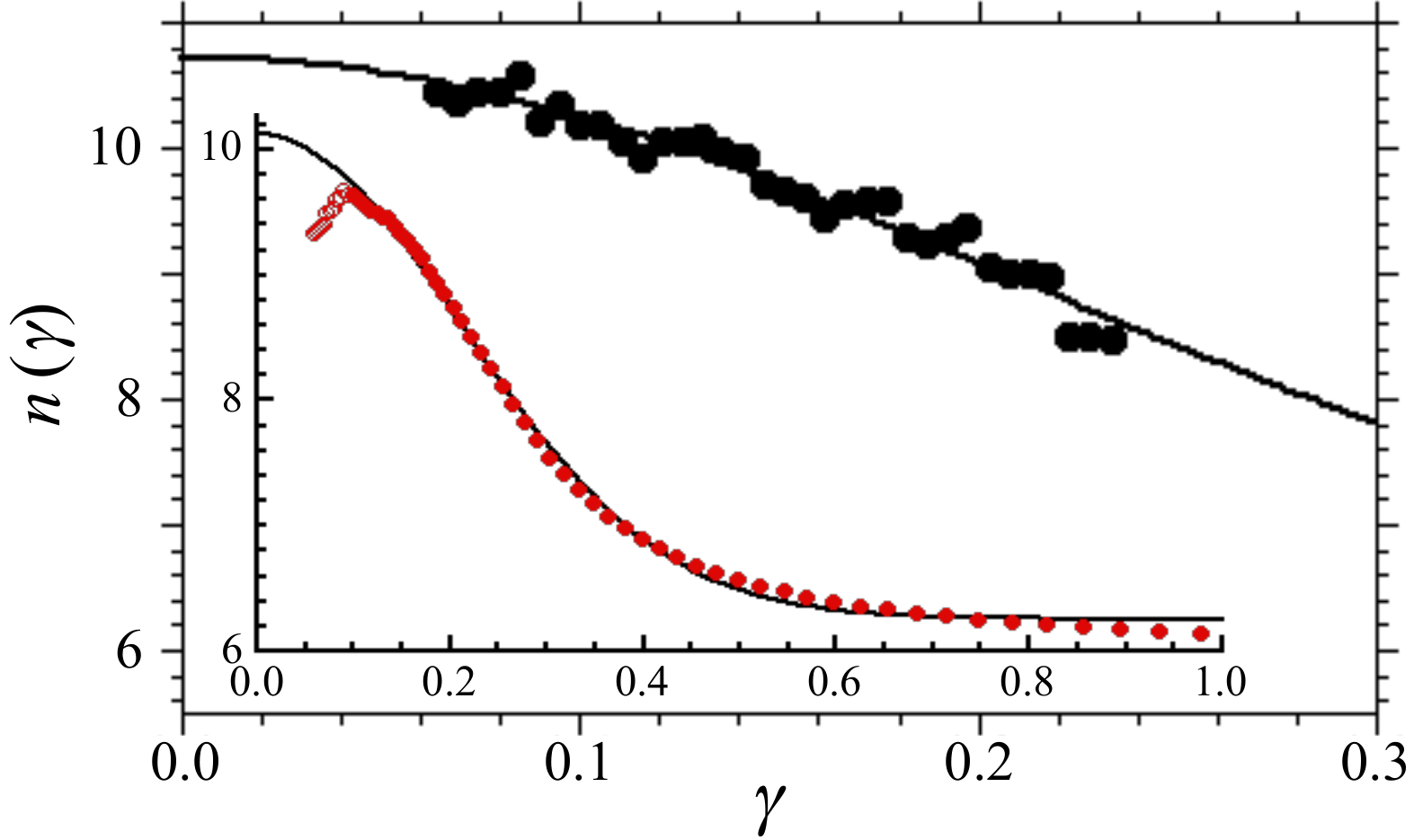}
\caption{(Color online) Mean number of long-lived neighbors, $n(\gamma)$, as a function of strain $\gamma$ obtained from (main figure) confocal microscopy and (inset) rheology for shear rate $\dot{\gamma}=0.0028$~s$^{-1}$, volume fraction $\phi=0.565$ and the large particles ($R_2=780$~nm).  The solid lines represent fits (Eq.~Ê\ref{eq:nb}), for which the first few data points (open circles) have been disregarded in the case of the rheological data.
}
\label{fig:nb}
\end{figure}

Based on $g(r)$ alone, long-lived neighbors cannot be distinguished from short-lived neighbors. Neighbors were considered long-lived if they remained neighbors during the lag time $\tau = 25 \Delta t = 8.0 \, \tau_{\text{B}}$ with the Brownian time $\tau_{\text{B}}=6\pi\eta_{\text{s}}R^3/(k_{\text{B}}T) = 5.7$~s, i.e.~if they were never further apart than $r=2.6R$ while the strain was increased from $\gamma - \dot{\gamma}\tau/2$ to $\gamma + \dot{\gamma}\tau/2$. This value of $\tau$ was chosen because it is the longest experimentally accessible lag time before the particles leave the field of view, but nevertheless short compared to the imposed time scale, i.e.~the inverse shear rate, $\dot{\gamma}\tau \ll 1$.
With this definition of `long-lived', we find the mean number of long-lived neighbors $n(\gamma)$ to decrease with strain $\gamma$ (Fig.~\ref{fig:nb}). Since the imposed deformation $\Delta \gamma = \dot{\gamma}\tau$ during the fixed time interval $\tau$ is independent of the accumulated strain $\gamma$, the decrease of $n(\gamma)$ with $\gamma$ is only due to the shear-enhanced mobility \cite{Koumakis2012, mutch2013, Koumakis2016, Sentjabrskaja2014, Zausch2008, Laurati2012} and not the increasing (affine) deformation $\gamma$.

The observed decrease of $n(\gamma)$ can be described by a super-exponential decay from an initial, $n_0$, to a final, $n_\infty$, value,
\begin{equation}
n(\gamma) = \left (n_0{-}n_\infty \right ) {\text{e}}^{-(\gamma/\xi)^2} + n_\infty
\label{eq:nb}
\end{equation}
with the characteristic decay parameter $\xi$ (Fig.~\ref{fig:nb}).
This appears reasonable as a neighbor is more likely to leave once other neighbors have left the cage and cage rearrangements have occurred. 
Super-exponential dependencies have also been observed in related situations \cite{Cipelletti2000, Evenson2015, Bouchaud2001}.
Once the system has become fluid-like, a finite $n_\infty = 6$ is expected due to the steady-state hydrodynamic flow and its local structure \cite{Brady1997}. This flow pushes the six neighbors in compression direction towards the particle and thus they remain for a long time.
A fit yields $n_0=10.7$.
Together with $n_{\mathrm{tot}}=11.6$, this implies that initially on average only about one neighbor leaves during the lag time $\tau$, while the remaining $n_0$ neighbors are `long-lived'. This appears reasonable given the large $\phi$.
Neighbors become considerably shorter lived as the system starts to flow under shear \cite{Koumakis2012,mutch2013, Koumakis2016,Sentjabrskaja2014, Sentjabrskaja2013}.
In the steady state ($\gamma \to \infty$) about six neighbors leave during $\tau$. 
The characteristic strain during which the steady state is approached was found to be $\xi=0.31$. This is consistent with rheological data obtained previously \cite{Koumakis2012, mutch2013, Koumakis2016, Sentjabrskaja2014, Zausch2008, Amann2013, Laurati2012, Varnik2004} and with new data shown below.
Thus the parameterization, Eq.~\ref{eq:nb}, is supported by experimental evidence.

\begin{figure}
\centering
\includegraphics[width=0.8\columnwidth]{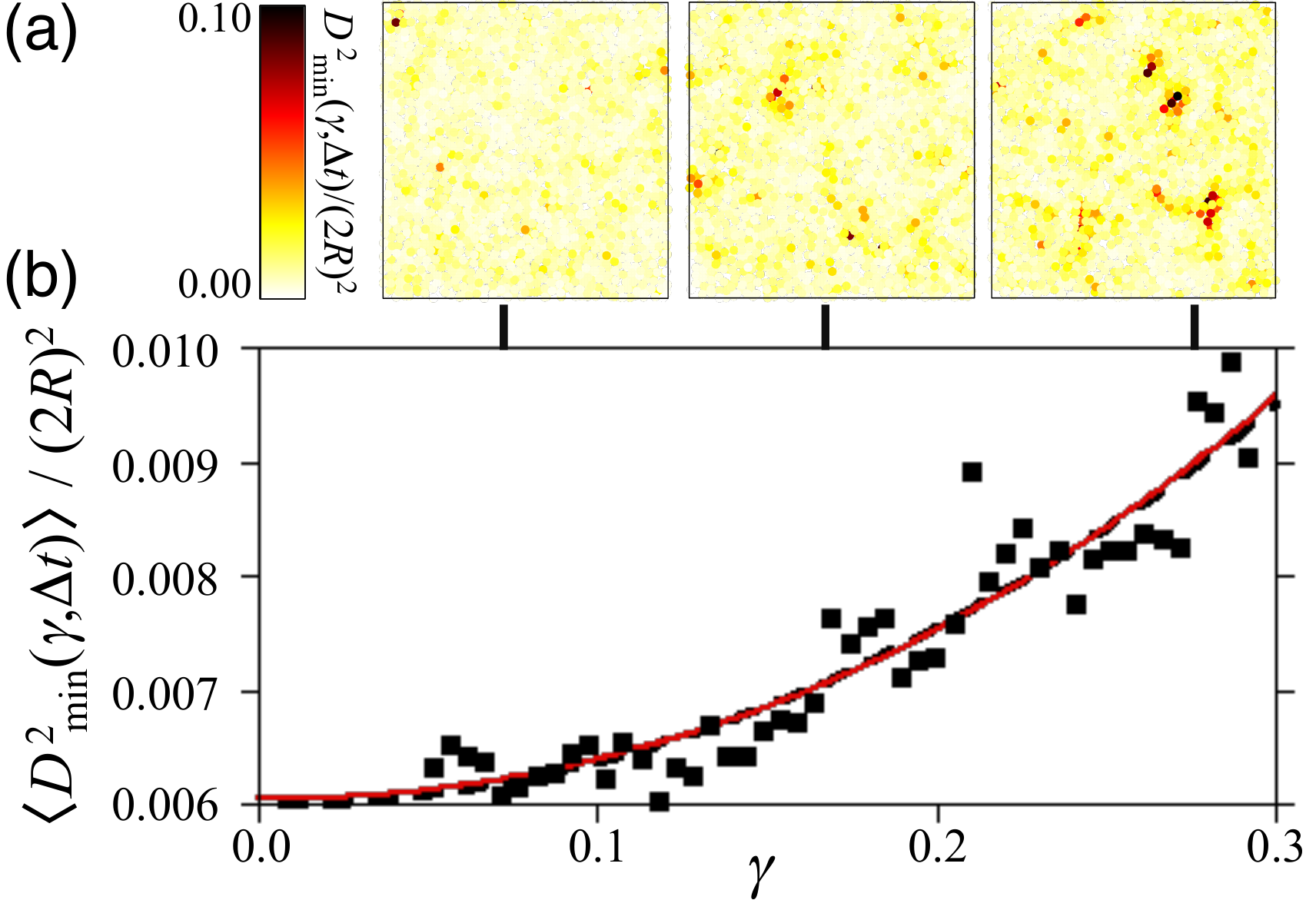}
\caption{(Color online) (a) Slices in the velocity-vorticity plane rendered from confocal microscopy images 
at strains $\gamma = 0.072$, $0.164$ and $0.277$ (left to right). The squared nonaffine displacement of each particle during lag time $\Delta t$, quantified by $D_{\text{min}}^2(\gamma, \Delta t)/(2R)^2$, is indicated by the color of the spheres. (b) Ensemble-average $\langle D^{2}_{\text{min}}(\gamma, \Delta t) \rangle/(2R)^2$ versus $\gamma$. The solid and dashed lines represent (almost identical) fits with cubic and quadratic order, respectively
(Eq.~\ref{eq:uNA2}).
}
\label{fig:D2}
\end{figure}

To quantify nonaffine rearrangements on a local (cage) level, we followed an established formalism \cite{falk_langer,chikkadi2012,Varnik2004}. 
First, the actual displacements of a particle's neighbors during a lag time $\delta t$ are determined.
Second, one calculates the notional displacements of the same neighbors during the same $\delta t$ in a fictitious affine rearrangement that is based on a local strain $\gamma(\vec{r})$.
Then the squared difference $D^2(\delta t)$ between these two displacements is calculated and $\gamma(\vec{r})$ chosen to minimize $D^2(\delta t)$.
This minimum value, $D_{\text{min}}^2(\delta t)$, characterizes the smallest deviation of the neighbors' displacements from affine displacements. Hence it quantifies the nonaffine rearrangements on a single-particle level. The lag time $\delta t$ should be short compared to $t=\gamma/\dot{\gamma}$. Following previous work \cite{chikkadi2012}, we take $\delta t = \Delta t$, when the motion is expected to be maximally correlated, i.e.~non-Gaussian. For each pair of image stacks at $\gamma$, $D_{\text{min}}^2(\gamma, \Delta t)$ is determined and its ensemble average, $\langle D_{\text{min}}^2(\gamma, \Delta t) \rangle$, calculated. It is found to increase with $\gamma$ (Fig.~\ref{fig:D2}). This increase in non-affine rearrangements is consistent with the decrease of the number of long-lived neighbors, $n(\gamma)$ (Fig.~\ref{fig:nb}).

The $\langle D_{\text{min}}^2(\gamma, \Delta t) \rangle$ characterizes the mean squared local nonaffine displacement, $u_{\text{NA}}^2(\gamma) \sim \langle D_{\text{min}}^2(\gamma, \Delta t) \rangle/(2R)^2$. For glasses, nonaffine displacements are about $20\%$ of affine displacements \cite{falk_langer, Denisov} which, by definition, are $\gamma$. For small $\gamma$ thus $u_{\text{NA}} \simeq b \gamma$ with $b \approx 0.2$. At larger $\gamma$, higher-order terms might be important; $u_{\text{NA}} \simeq b\gamma+c\gamma^{2}$ with a phenomenological coefficient $c$. Thus, to cubic order
\begin{equation}
u_{\text{NA}}^{2}(\gamma) \simeq a+b^{2}\gamma^{2}+2bc\gamma^{3} \; ,
\label{eq:uNA2}
\end{equation}
where the offset $a$ accounts for noise arising from instabilities of the setup and uncertainties in the particle location. This yields a good fit (Fig.~\ref{fig:D2}) with $a=0.0060$,
$b=0.18$ (comparable to the expected value of about $0.2$ \cite{Denisov})
and $c=0.058$.
The small value of $c$ suggests $c \approx 0$. A fit with $c=0$ is almost identical and yields $a=0.0060$ and $b=0.20$ (Fig.~\ref{fig:D2}). For the applied strain $\gamma < 0.3$, therefore, a linear approximation seems sufficient in the present case.

\begin{figure}
\centering
\includegraphics[width=0.768\columnwidth]{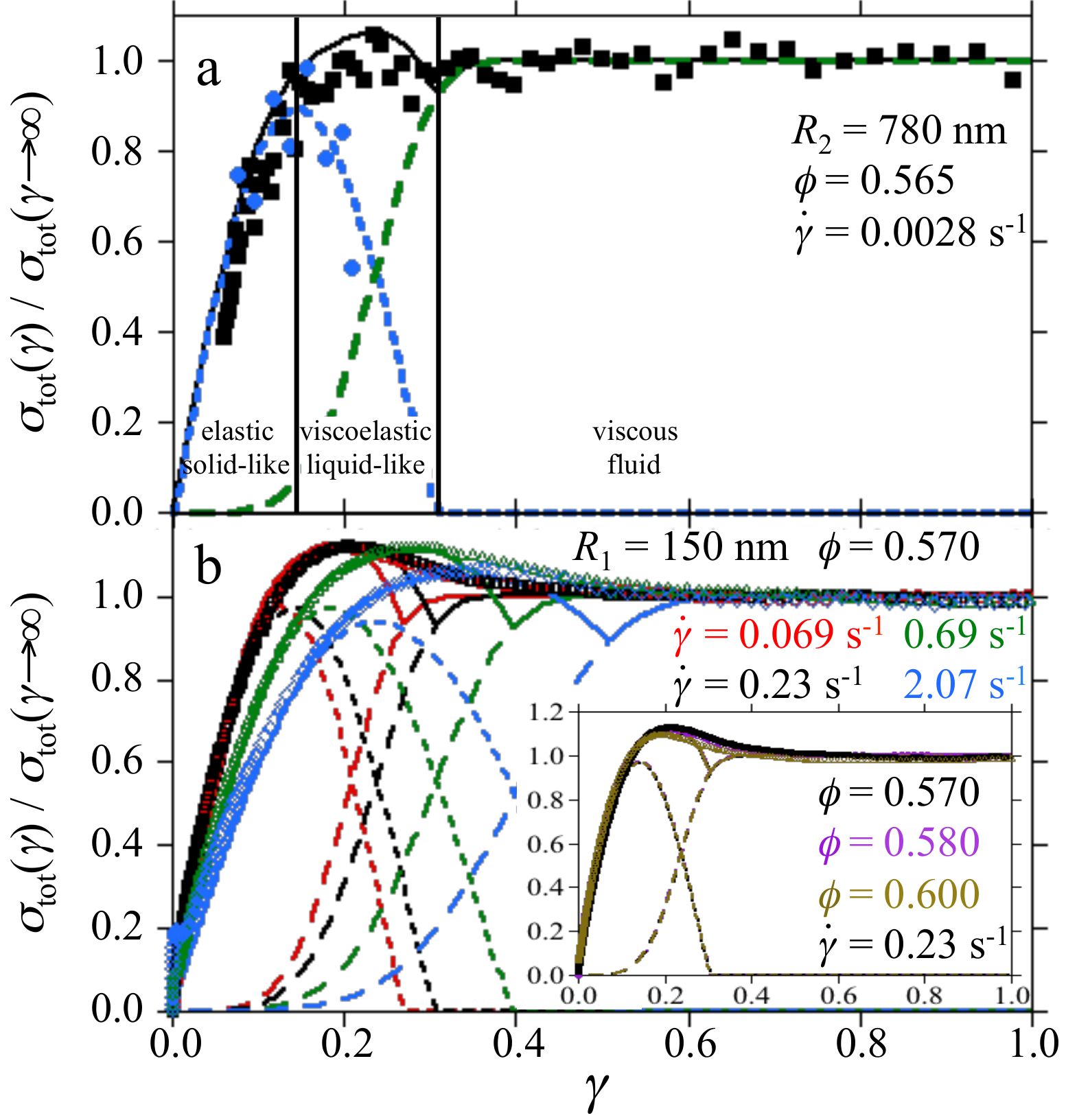}
\caption{(Color online) Normalized transient total stress $\sigma_{\text{tot}}(\gamma)/\sigma_{\text{tot}}(\gamma{\to}\infty)$ as a function of strain $\gamma = \dot{\gamma}t$ during start-up for (a) large particles and (b) small particles with different shear rates $\dot{\gamma}$ (main figure) and volume fractions $\phi$ (inset) with $\dot{\gamma}$ and $\phi$ as indicated \cite{xi}. Solid lines represent fits, dotted and dashed lines the elastic, $\sigma_{\text{el}}$ (Eq.~\ref{eq:sel}), and viscous, $\sigma_{\text{visc}}$ (Eq.~\ref{eq:svisc}), contributions, respectively. Blue circles in (a) indicate  $\sigma_{\text{el}}$ as determined from the individual $n(\gamma_i)$ (data points in Fig.~\ref{fig:nb}) without using Eq.~\ref{eq:nb} (each symbol represents the average of four data points).  The different regimes are indicated in (a).
}
\label{fig:sigma}
\end{figure}

For the same sample, the rheological response to the application of a constant strain rate $\dot{\gamma}$, i.e.~$\sigma_{\text{tot}}(\gamma)$, has been measured (Fig.~\ref{fig:sigma}a).
This will be compared to the transient total stress $\sigma_{\text{tot}}(\gamma)$ as calculated based on the microscopic parameters $n(\gamma)$ and $u_{\text{NA}}^2(\gamma)$.
 
According to the Born-Huang theory of lattice dynamics extended for an isotropic random distribution of particles as in a glass, the affine part of the shear modulus reads~\cite{Born1954, Zaccone2011a}
\begin{equation}
G_{\text{A}}(\gamma)=\frac{1}{5\pi}\frac{\kappa\phi}{R} \, n(\gamma)   \;  ,
\label{eq:GA}
\end{equation}
with the effective (entropic) spring constant of a bond $\kappa=[{\text{d}}^{2}V_{\mathrm{eff}}(r)/{\text{d}}r^{2}]_{r=r_{\text{m}}}$. It is linked to the minimum of the pair potential of mean force $V_{\mathrm{eff}}(r)$, located at $r=r_{\text{m}}$, and thus to the first maximum of the radial distribution function, $g(r)$, since $V_{\text{eff}}(r)=-k_{\text{B}}T \ln g(r)$~\cite{Hansen2005}. The experimental $g(r)$ yields $\kappa = 50.2 \,k_{\text{B}}T/R^2$ \cite{SM}, which is consistent with theoretical predictions \cite{Schweizer2003}.

Based on a third-order expansion of the nonaffine part of the elastic free energy density, $F_{\text{NA}}$~\cite{Alexander98}, the nonaffine part of the shear modulus, $G_{\text{NA}}=\partial^{2} F_{\text{NA}}/\partial \gamma^{2}$, is given to first order by
\begin{equation}
G_{\text{NA}}(\gamma)= \frac{1}{5\pi}\frac{\kappa\phi}{R}~(6+c\gamma) \; .
\label{eq:GNA}
\end{equation}
The term linear in $\gamma$ takes into account the nonlinear behavior, quantified by $u_{\text{NA}}^2(\gamma)$ (Eq.~\ref{eq:uNA2}). In our sample, the absence of higher-order terms in $u_{\text{NA}}^2(\gamma)$ implies a $\gamma$ independent $G_{\text{NA}}$. Nevertheless, it constitutes an important (constant) negative contribution to the total shear modulus. It also implies that the departure from linearity of the shear modulus is dominated by the loss of long-lived neighbors, i.e.~the decrease of $G_{\text{A}}$.

The shear moduli are linked to the corresponding affine, $F_{\text{A}}$, nonaffine, $F_{\text{NA}}$, and total, $F_{\text{el}}$, elastic free energies; $F_{\text{el}}(\gamma)=F_{\text{A}}(\gamma)-F_{\text{NA}}(\gamma) = \frac{1}{2}(G_{\text{A}}{-}G_{\text{NA}})\gamma^2$.
The elastic (reversible) stress $\sigma_{\text{el}}=\partial F_{\text{el}}(\gamma)/\partial \gamma$ hence is
\begin{eqnarray}
\sigma_{\text{el}}(\gamma) &=& \frac{\kappa\phi}{10\pi R} \left \{ 2 \left (n_0{-}6 \right ) \gamma \left ( 1 - \left ( \frac{\gamma}{\xi} \right )^2 \right ) e^{-\left (\frac{\gamma}{\xi} \right )^2}  -  3 c \gamma^2  \right \} \nonumber
\label{eq:sel}
\setcounter{equation}{6}
\end{eqnarray}
For $c \ne 0$, this implies $\sigma_{\text{el}}(\gamma{\to}\infty) \to -\infty$ which is unphysical. To avoid introducing free parameters, $\sigma_{\text{el}}(\gamma)$ is assumed to attain a constant value beyond the characteristic strain $\xi$ of the exponential decay; $\sigma_{\text{el}}(\gamma{\ge}\xi) = \sigma_{\text{el}}(\xi)$, which here vanishes since $c=0$.

For small strains $\gamma$, $G=[\partial \sigma_{\text{tot}}(\gamma)/\partial \gamma]_{\gamma{\to}0} \approx [\partial \sigma_{\text{el}}(\gamma)/\partial \gamma]_{\gamma{\to}0}=(\kappa\phi / 5 \pi R)(n_0{-}6)$, as previously calculated \cite{Zaccone2011, Zaccone2011a}. Hence, from the initial slope we can obtain $\kappa$, here $\kappa = 49.7 \,k_{\text{B}}T/R^2$ in agreement with the value based on $g(r)$ from confocal microscopy.

The viscous (dissipative) stress, $\sigma_{\text{visc}}$, is due to internal friction. For a linear viscoelastic solid, just as for a Maxwellian fluid, under start-up shear, $\sigma_{\text{visc}}=\dot{\gamma} \eta \{ 1-\exp{(-\gamma/\dot{\gamma}\tau_{\text{v}})}\}$~\cite{Oswald}. In glassy systems, however, the complex energy landscape leads to non-exponential behavior, typically described by \cite{Bouchaud2008}
\begin{equation}
\sigma_{\text{visc}}(\gamma)=\dot{\gamma} \eta \left \{1{-}{\text{e}}^{-\left (\gamma/\dot{\gamma}\tau_{\text{v}} \right )^\beta}\right \} \; .
\label{eq:svisc}
\end{equation}
The viscosity of the suspension, $\eta$, is determined from the steady state, $\sigma_{\text{tot}}(\gamma{\to}\infty)\approx \sigma_{\text{visc}}(\gamma{\to}\infty)=\dot{\gamma}\eta$, i.e.~the flow curve. Here, $\eta = 2.6$~Pa\,s (Fig.~\ref{fig:sigma}a). Furthermore, $\eta$ is linked to $G$ through $\eta = \eta_{\text{s}} \exp{(G V^\ast /k_{\text{B}}T)}$, where $V^\ast$ is an activation volume expected to be of the order of a particle volume \cite{Leporini, Dyre1998}.
This is indeed found; $V^\ast=0.20\, (4\pi/3)R_2^3$.
The viscous time scale $\tau_{\text{v}}$ \cite{Cipelletti, DelGado, Barrat, Sastry} can be estimated through Maxwell's expression \cite{Dyre1998}, $\tau_{\text{v}} = \eta / G = 36\,{\text{s}} = 6.3\,\tau_{\text{B}}$,
and hence can be determined from the initial slope and steady-state value of $\sigma_{\mathrm{tot}}$. 
For the stretching exponent $\beta$, typically $1.5 \lesssim \beta \lesssim 1.8$ is found for quiescent soft glasses and gels~\cite{Cipelletti, DelGado, Barrat} and significantly larger values in the hydrodynamic limit, $\beta>2$~\cite{Cipelletti,Sastry}. Since the present system is in the hydrodynamic limit and subject to shear, a value considerably larger than $2$ is expected. This is also consistent with the reported superdiffusive dynamics \cite{mutch2013, Zausch2008, Laurati2012}.

Finally, the transient total stress $\sigma_{\text{tot}}(\gamma)=\sigma_{\text{el}}(\gamma) + \sigma_{\text{visc}}(\gamma)$~\cite{Landau}.
A fit with $\tau_{\text{v}}$ and $\beta$ as the only free parameters, for which nevertheless estimates exist, describes the experimental data very well (Fig.~\ref{fig:sigma}a). Only at the transition from elastic to viscous behavior a spurious dip occurs at $\gamma \approx \xi$ due to the abrupt transition to the constant value $\sigma_{\text{el}}(\xi)$, as anticipated and discussed above (Eq.~\ref{eq:sel}). We refrain from a more detailed and `smoother' modeling of this transition to avoid increasing the number of adjustable parameters. The values of the fit parameters are consistent with expectations; $\tau_{\text{v}} = 89\,{\text{s}} \approx 16\,\tau_{\text{B}}$ is similar to the predicted value of about $6.3\,\tau_{\text{B}}$, and $\beta = 4.6$ is considerably larger than 2, as expected. 

The elastic stress $\sigma_{\text{el}}$ was also determined based on the individual $n(\gamma_i)$ (data points in Fig.~\ref{fig:nb}), i.e.~without using the empirical dependence of the mean number of long-lived neighbors $n$ on strain $\gamma$ (Eq.~\ref{eq:nb}).
The agreement with the calculation using the empirical relation (Fig.~\ref{fig:sigma}a) indicates that it does not introduce artifacts.
Furthermore, from the rheology data, i.e.~the total stress $\sigma_{\text{tot}}(\gamma)$, the mean number of long-lived neighbors, $n(\gamma)$, was calculated using the same relationships but in inverse order (Fig.~\ref{fig:nb}, inset). 
A fit based on Eq.~\ref{eq:nb} yielded similar parameters,  $\xi=0.30$, $n_0=10.1$ and $n_\infty=6.2$.
This supports the consistency of the model and the empirical dependence of $n$ on $\gamma$ (Eq.~\ref{eq:nb}).

The dependencies on $\phi$ and $\dot{\gamma}$ were investigated using smaller particles.
The normalized $\sigma_{\text{tot}}(\gamma)/\sigma_{\text{tot}}(\gamma{\to}\infty)$ are almost independent of $\phi$ within the range investigated (Fig.~\ref{fig:sigma}b, inset), whereas a strong dependence on $\dot{\gamma}$ is observed (Fig.~\ref{fig:sigma}b), as previously reported \cite{Laurati2012,mutch2013, Koumakis2016, Koumakis2012, Zaccone2014, Sentjabrskaja2014}.
The small particles result in a strong rheological signal, but preclude the use of confocal microscopy to obtain the parameters describing the long-lived neighbors and the nonaffine displacements. 
Thus, we assume that $n_0=10.7$ and $c=0$ for all investigated $\phi$ and $\dot{\gamma}$, while one value of $\xi$ was fitted for each additional $\dot{\gamma}$. Apart from these three values for $\xi(\dot{\gamma})$ (bold in Tab.~SM-2), all other values can be deduced based on the relationships introduced above, the observation that $\sigma_{\text{tot}}(\gamma)/\sigma_{\text{tot}}(\gamma{\to}\infty)$ is almost independent of $\phi$, and the values determined for the large spheres. (Details of the fitting procedure and the fitted values are given in the SM \cite{SM}.)
The fits to $\sigma_{\text{tot}}(\gamma)$ show very good agreement with the data, again, except for the expected spurious dip (Figs.~\ref{fig:sigma}b, SM-2).
We found that $\xi$ increases with $\dot{\gamma}$, as previously predicted \cite{Zaccone2014}. This is attributed to the increasing importance of shear compared to Brownian motion and hence the dominance of affine motions which result in neighbors remaining neighbors for larger strains.
Furthermore, $\tau_{\text{v}}(\dot{\gamma}) \sim \eta(\dot{\gamma})/\kappa(\dot{\gamma})$ decreases with $\dot{\gamma}$, as expected for an increasingly fluidized system with the diffusion coefficient $D(\dot{\gamma}) \sim \dot{\gamma}^{0.8}$ \cite{Besseling2007} or $\dot{\gamma}^{1}$ \cite{Miyazaki2006}. Since $\tau_{\text{v}}$ represents a relaxation time under shear, $\tau_{\text{v}} \sim D^{-1} \sim \dot{\gamma}^{-0.8}$ or $\dot{\gamma}^{-1}$. Our data indicate $\tau_{\text{v}} \sim \dot{\gamma}^{-0.80}$, in agreement with one of the previous findings \cite{Besseling2007}. 

In all cases, the response shows three distinct regimes (Fig.~\ref{fig:sigma}a). At very low strains, the elastic response dominates ($\sigma_{\text{el}} \gg \sigma_{\text{visc}}$); the sample behaves solid-like. Upon increasing the strain, the number of long-lived neighbors $n(\gamma)$ decreases (Fig.~\ref{fig:nb}) and cages break. Concomitantly, the nonaffine displacements $u_{\text{NA}}^2(\gamma)$ significantly increase (Fig.~\ref{fig:D2}). The system yields and the response is dominated by nonaffine rather than affine motion. Hence the nonaffine elastic free energy becomes important and leads to a significant negative contribution to $F_{\text{el}}$ which hence vanishes, as does $G$. Then, in the second viscoelastic liquid-like regime, the elastic component decreases and the increasing viscous dissipative component becomes dominant. Once the elastic contribution vanishes, the stress is entirely viscous and the glass turns into a viscous fluid.


To conclude, we related the macroscopic transient rheological response of glasses to the shear-induced microscopic structural evolution. The \textit{long-lived} nearest neighbors forming the cage were identified to be the crucial structural feature with their number $n(\gamma)$ decreasing under shear. At the same time, \textit{nonaffine} rearrangements of the neighbors, quantified by the mean squared local nonaffine displacement $u_{\text{NA}}^2(\gamma)$, increase. Both microscopic parameters are experimentally determined and form the basis for the analytical calculation of the rheological response. The affine and nonaffine contributions to the elastic free energy and its derivative, the elastic stress, are supplemented by the viscous stress to give the total stress. The predictions for the transient total stress agree with experimental rheological data for different volume fractions $\phi$ and shear rates $\dot{\gamma}$. Three regimes are identified that are dominated by the affine, nonaffine and viscous contributions, respectively. The agreement supports the proposed framework and its quantitative link between the macroscopic rheological response and the microscopic structural and dynamical evolution.


\begin{acknowledgments}

We thank A.B.~Schofield for the particles and acknowledge support from the Deutsche Forschungsgemeinschaft (DFG) through the research unit FOR 1394, project P2.

\end{acknowledgments}


\end{document}